\documentclass[conference]{IEEEtran}
\usepackage{cite}
\usepackage{amsmath,amssymb,amsfonts}
\usepackage{multirow}
\usepackage{algorithmic}
\usepackage{graphicx}
\usepackage{textcomp}
\usepackage{xcolor}
\usepackage{amsmath}
\usepackage{url}
\usepackage{balance}
\usepackage{mathtools}

\DeclarePairedDelimiter\ket{\lvert}{\rangle}

\begin{document}

\title{OpenSurgery for Topological Assemblies}

\author{
\IEEEauthorblockN{Alexandru Paler}
\IEEEauthorblockA{\textit{Johannes Kepler University}, 4040 Linz, Austria}
\and
\IEEEauthorblockN{Austin G. Fowler}
\IEEEauthorblockA{\textit{Google Inc.}, Santa Barbara, 93117 CA, USA}
}

\maketitle

\begin{abstract}
Surface quantum error-correcting codes are the leading proposal for fault-tolerance within quantum computers. We present OpenSurgery, a scalable tool for the preparation of circuits protected by the surface code operated through lattice surgery. Lattice surgery is considered a resource efficient method to implement surface code computations. Resource efficiency refers to the number of physical qubits and the time necessary for executing a quantum computation. OpenSurgery is a first step towards methods that aid quantum algorithm design informed by the realities of the hardware architectures. OpenSurgery can: 1) lay out arbitrary quantum circuits, 2) estimate the quantum resources used for their execution, 3) visualise the resulting 3D topological assemblies. Source code is available at \url{http://www.github.com/alexandrupaler/opensurgery}.
\end{abstract}

\section{Introduction}

Practical quantum computing will be almost impossible in the absence of quantum error correcting codes (QECC), because the quantum hardware is not as reliable as the classical hardware. QECCs need to handle comparatively high error rates, and at the same time the QECCs should introduce low hardware (physical qubits) and time overheads. An ideal QECC should have a straightforward structure, such that they are easy to implement in hardware. One of the realistic QECC choices that fulfils most of the previous conditions is the surface QECC, and major quantum computer proposals are using it to implement reliable quantum computations. Surface code protected computations can be implemented by braiding \cite{fowler2012surface}, lattice surgery\cite{horsman2012surface,litinski2018game}, and twists \cite{brown2017poking}.

Lattice surgery seems to be the most resource efficient option\cite{litinski2018game}, because its overheads are for most of the usual computations lower than when using braiding \cite{paler2017synthesis}. In the following, we introduce OpenSurgery, the first compiler of arbitrary quantum circuits to lattice surgery structures. We describe the design of OpenSurgery, its features and performance. Finally, we formulate future work.

\begin{figure}[t!]
    \centering
    \includegraphics[width=0.9\columnwidth]{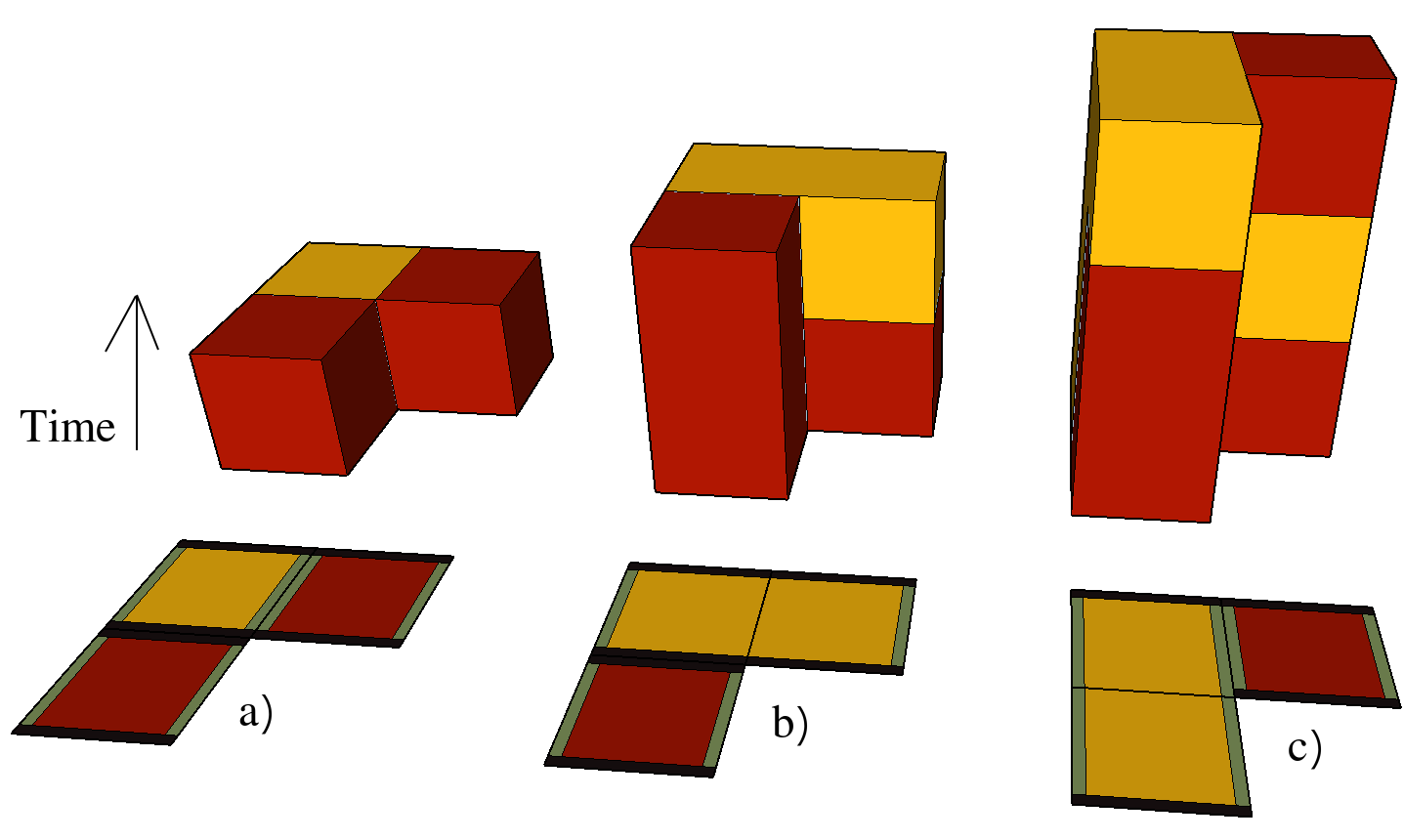}
    \caption{Lattice surgery CNOT using three patches. The yellow patch is an ancilla and intermediates between the red patches. Merge and split operations are performed along patch boundaries (green and black). Operating a patch for a given time is abstracted as a cuboid. The back row illustrates the topological assembly resulting after the two merge and splits. The front row indicates how boundaries are operated: a) identity operation; b) merge and split along the green boundary; c) merge and split along the black boundary.}
    \label{fig:1}
\end{figure}

This section includes a high level overview of the absolutely necessary concepts for introducing OpenSurgery. A detailed introduction to surface codes and lattice surgery is available in \cite{fowler2012surface} and \cite{horsman2012surface}. Lattice surgery uses \emph{patches} (e.g. Fig.~\ref{fig:1}) to represent logical qubits. Each patch is defined over a rectangular region of physical qubits arranged in a two-dimensional lattice. Patch dimensions indicate the distance of the surface code, and this influences the number of physical qubits as well as the time necessary for the error-correction. Each patch has two boundary types, and opposite boundaries are of the same type (e.g. in Fig.~\ref{fig:1} there are green and black boundaries). Quantum circuit gates are implemented by sequences of \emph{merge} and \emph{split} operations along patch boundaries. The types of boundaries used influences the implemented gate functionality.

Compiling circuits for lattice surgery is a multiple step procedure. First, circuit qubits are mapped to patches. Second, gates are decomposed into split and merge operations. Third, the resulting 3D structures representing qubits and operations are arranged in a space-time volume according to well define rules. The compilation result is called a \emph{topological assembly} and it is a set of cuboids (e.g. Fig.~\ref{fig:1}). Two dimensions of the cuboids are the code distance, and the third dimension the duration for which error-correction is applied.

The three dimensional bounding box of the assembly is called the \emph{equivalent volume} and it is a measure of the resources needed to execute the encoded computation (physical qubits and the time). The equivalent volume is obtained by multiplying the area of the layout (e.g. 4 in Fig.~\ref{fig:1} because we consider the footprint being a square of four patches and not only three) with the number of steps (e.g. three). The assembly in Fig.~\ref{fig:1} has a volume of 12.

\section{Methods}

\begin{figure}[t!]
    \centering
    \includegraphics[width=0.9\columnwidth]{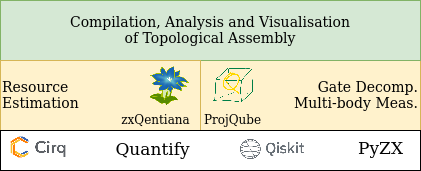}
    \caption{The architecture of OpenSurgery. Circuits from different frameworks can be used in OpenSurgery. The resource estimation engine is based on zxQentiana, and the compiler of multi-body measurements is ProjQube.}
    \label{fig:arch}
\end{figure}

OpenSurgery is a scalable research-grade software tool with a modular architecture. Fig.~\ref{fig:arch} includes the major components and their interaction. The tool has an interface to IBM Qiskit by the usage of the OpenQASM circuit description language. Other circuit formats, such as Quipper, can be imported into OpenSurgery through PyZX. The OpenQASM interface can be used also for importing from Google Cirq, but for such circuits the preferred method is over QUANTIFY\footnote{\url{https://www.github.com/quantumresources/quantify}} which is natively built on top of Cirq.

OpenSurgery generates assemblies starting from Clifford+T circuits, as well as for  circuits including arbitrary rotation gates. The latter gates are decomposed by a Solovay-Kitaev algorithm implementation \footnote{Python version based on \url{https://github.com/cryptogoth/skc-python}} into Clifford+T gates. This can be done more efficiently using modern techniques. However, our focus here is on laying out a Clifford+T circuit, not achieving the smallest circuit. Internally, the Clifford+T circuits are then transformed into sequences of multi-body (multi-qubit) measurements which are implemented by merges and splits. 

For lattice surgery surface code computation, single qubit gates can be implemented: a) directly on the patch (e.g. the Hadamard, or the S gate); or b) by merges and splits. For example, the T gate requires merges and splits, because it is implemented by teleportation using a CNOT gate. The CNOT, an example of a multi-qubit operation, is a sequence of two circuits implementing multi-body measurements (e.g. Fig.~\ref{fig:1}). The functionality of OpenSurgery is controlled through a list of instructions, which are described in Sec.~\ref{sec:instr}.

\subsection{Automatic Layout and Routing}
\label{sec:routing}

\begin{figure}
    \centering
    \includegraphics[width=0.7\columnwidth]{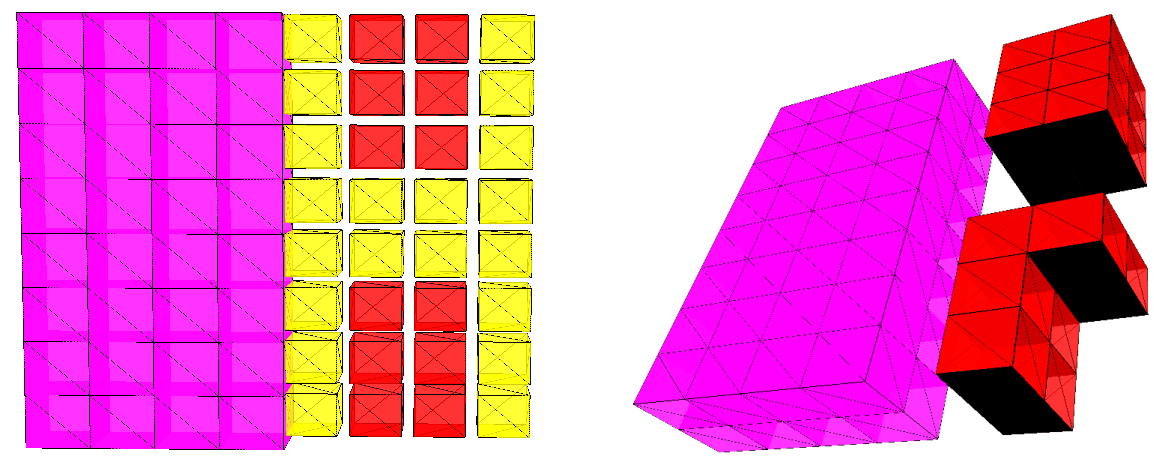}
    \caption{Layout of a topological assembly. (a) The left region is for distillations (magenta), data patches are in rows (red), ancilla patches (yellow) are between the data patches to intermediate multi-body measurements. (b) Patches were placed in two layers. The ancilla are not used, and the distillation procedure is being executed. Execution time flows from image background to foreground.}
    \label{fig:2}
\end{figure}

The layout of an assembly refers to the footprint of the assembly (the two dimensions for hardware). There are multiple ways of arranging a layout into regions. OpenSurgery is very flexible, and discretises the layout into patches (tiles) of one of four region types: 1) distilleries; 2) queues of distilled states \cite{paler2019clifford}; 3) logical qubit patches; 4) ancillae patches. For example, in Fig.~\ref{fig:2} the magenta region is for a single distillation. The queues are where distilled states are stored such that T gates can be implemented without waiting for the distillation to finish. This may happen if the consumption rate of distilled states is not constant.

Circuit qubits are mapped to logical qubit patches. Quantum gates between the patches are implemented by using the ancillae patches. There are three parameters which influence the resource efficiency of a layout: a) the number of distillations executed in parallel; b) the total number of logical qubits; c) the maximum routing overhead. The latter is the ratio between the number of ancillae and the number of logical qubit patches. All three parameters can be configured in OpenSurgery. The most resource efficient layout includes a distillery for a single distillation at a time.

Minimising the routing overhead is desireable, too. This is a function of: a) the average number of logical qubits involved in gate operations; b) how qubits were mapped to patches (e.g. it may be more efficient to cluster often interacting patches). OpenSurgery uses an ancilla arrangement that supports all-to-all connectivity between logical qubit patches. This  simplifies the realisation of merges and splits, but it is not necessarily the most resource efficient. Connecting data patches is a \emph{routing problem}: finding the shortest path formed of ancillas between two data patches. OpenSurgery computes the shortest paths between pairs of data patches using the A* algorithm.

\subsection{OpenSurgery Instruction Set}
\label{sec:instr}

The assembly is generated by processing a sequence of multibody measurements. These are obtained from arbitrary Clifford+T circuit gate lists by using the \emph{projCube}\footnote{\url{https://www.github.com/quantumresource/projqube}} software.

OpenSurgery includes instructions to implement the steps necessary to achieve the multibody measurements and the quantum gates. As a result, a topological assembly can be seen as the result of the OpenSurgery compiler responding to a sequence of instructions for the reservation, manipulation and consumption of resources (e.g. distilled states, ancilla patches). The instruction set will be adapted and improved in future versions of OpenSurgery. 

For single qubit gates which are implementable without ancillae in the surface code, the instructions resemble the names of the gates: \texttt{V=$\sqrt{X}$},  \texttt{S=$\sqrt{Z}$} and \texttt{H}. However, instructions are more than names for quantum gates. For example, for a T gate, a distilled A state is necessary. The instruction $NEED_A$ instructs the queue of distilled states to: a) either provide a distilled state, or b) to request the distillery to distil a state. This procedure was described in \cite{paler2019clifford}. For the moment, OpenSurgery uses layouts that include: a) a queue of capacity one for distilled states, b) a distillery distilling a single state a time. Consequently, the \texttt{NEED\_A} instruction is blocking, in the sense that, once called, no other instructions are executed until a distilled state is available. Once that is the case, to implement the T gate, the next instruction is \texttt{MZZ A [qub]} for ZZ-measuring simultaneously the patch of the distilled A state and the patch of the logical qubit $qub$. Finally, the \texttt{MX A} instruction tells OpenSurgery to measure in the $X$-basis the path of the distilled A state. T gate implementation is probabilistic, and OpenSurgery will automatically provision an $S$ gate instruction in case a state correction is needed.

\begin{figure*}[h!]
\centering
\includegraphics[width=0.9\textwidth]{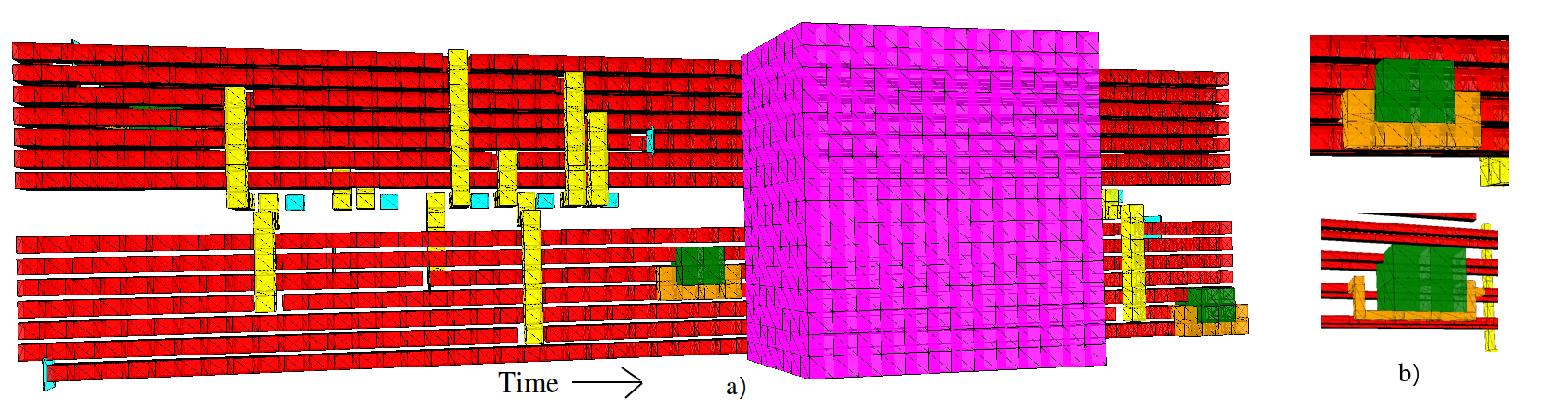}
\caption{Compiled assembly that includes a distillation. a) Yellow segments are multi-body measurements. The horizontal empty space between the red segments (logical qubits) is the ancillae region (cf. Fig.~\ref{fig:2}). Green boxes represent S gates, and orange segments are qubits re-routed for the S gate to be executed. b) The same S gate assembly but with a different cuboid scaling and distance between patches in order to ease visualisation.}
\label{fig:screen}
\end{figure*}

Another instruction is \texttt{INIT [qub]}, for initialising a qubit patch in the $\ket{0}$ state. The \texttt{ANCILLA [state]} instruction requests the initialisation of an ancilla patch in $\ket{0}$ or $\ket{+}$. The \texttt{MOVE patch1 patch2} instructs OpenSurgery to move the logical qubit state from patch1 to patch2. This can be performed by SWAP gates if along the computed path there are other occupied patches, or simply by splits and merges between all the patches on the path connecting patch1 and patch2. The latter situation is often the case when not all the ancillae patches are being used. Some instructions require one time step, while others (e.g. the S gate in Fig.~\ref{fig:screen}) take two or more steps. This is because of how the different operations are implemented by physical hardware instructions (e.g. circuit depth needed for a non-transversal gate).

\subsection{Resource Estimation}
\label{sec:res}

The volume of the compiled topological assembly can be used to estimate the total number of physical qubits and the time needed to executed the assembly. The estimation procedure follows \cite{fowler2012surface} and uses the zxQentiana\footnote{\url{https://www.github.com/quantumresource/zxQentiana}} software. Compared to \cite{fowler2012surface}, the procedure was improved wrt. to the number of parameters i.e. distillation box sizes, placement of distillations in multilevel arrangements, code distances, physical error-rates, target logical error-rate, automatic computation of number of levels of distillations required, as well the capability to recommend a minimum distance. The code assumes that two levels of magic state distillation are possible and that the distillation surface code distances are different from the distance of the logical qubits. Default distillation distances are 15 and 31. Default patch distance is 7, and it is increased linearly until the target logical error-rate is achieved.

To maintain the scale, the assembly is generated and visualised (Sec.~\ref{sec:3d}) with cuboids of size the approximate largest common divisor of all the distances existing in the computation. For example, if distances 15 and 7 are used, distance 15 cuboids are drawn with cuboids of distance 7.

\subsection{Correlation Surface Tracking}

Patch boundaries are initialised at predetermined positions. For example, X-type boundaries in the north and south of the patch, and Z-type boundaries in the east and west of the patch. For a shorter implementation of the computations, it is sometimes advantageous to \emph{rotate the patches}, meaning that the position of the boundaries is interchanged by low level quantum circuits that implement the surface code in a careful manner not to affect the distance of the error-correction. Patch rotations do not change the stored logical state. 

Bit information along the patch boundaries can be tracked. This information indicates the sign of the logical operator of the patch. A correlation surface is defined as the discrete collection of all the logical operators ordered in space and time. The term correlation surface originates from braided topological code diagrams and is related to the fact that logical gates are implemented in a measurement-based manner (i.e. the analogy is the so-called gflow\cite{perdrix2007generalized}). Correlation surface interactions can be extracted from the lattice surgery diagrams as stabilizer truth tables using the ICM formalism \cite{paler2018specification}. Correlation surfaces are also, more conveniently, called space-time stabilizers, because they relate the operators of the circuit inputs to the operators of the outputs.

OpenSurgery keeps track of how the boundaries are positioned on the patches. This allows the tracking of correlation surfaces/spacetime stabilizers. The tracker includes a certain degree of automation: it determines automatically if for a particular multi-body measurement the boundaries are oriented correctly and, otherwise, these are rotated. In a pair of neighbouring patches, by rotating one but not the other, it is possible, for example, to perform an XZ-measurement instead of only XX- or ZZ-measurements. The work of \cite{litinski2018game} also mentions a kind of rotation where a single boundary has both types at the same time (half of the boundary is of one type, the other half of a different type) -- constructing such boundaries will be supported in future versions of OpenSurgery.

\subsection{3D Visualisation}
\label{sec:3d}

OpenSurgery includes a complete visualisation of the generated topological assemblies. This can be used for debugging as well as for visually inspecting potential optimisations of the resulting assembly. A topological assembly is presented as a group of cuboids, each being colour coded (e.g. distillation cuboids have a distinct colour from logical qubit patches), holding the identification of the gate/measurement that they belong to, as well as the space and time coordinates. The core of the visualisation are JSON files storing the assembly information and custom developed WebGL classes that use three.js. The performance of the visualisation speed depends on the computer's hardware. The figures from this paper are screenshots of visualised assemblies.

To ease the visualisation of patches and their interaction, OpenSurgery can automatically scale the patches and introduce connecting elements between them. An example is Fig.~\ref{fig:screen}b.

The visualisation tool can filter cuboids. For example, it is possible to see only the output distilled states, or to check how often a space coordinate is used along the assemblies execution. Such filters are useful for automating future assembly analysis and optimisation methods.

\section{Results}

Previous works analysed the resource efficiency of various patch arrangements \cite{javadi2017optimized}, or investigated surgery implementations of distillation circuits \cite{fowler2012surface, holmes2019resource}. No work focused on the automatic compilation of arbitrary circuits which include distilleries, allowing at the same time the resource estimation and the 3D visualisation of the topological assemblies.

We analyse the scalability of OpenSurgery using random circuits (available in the project repository). Two scenarios are executed: 1) constant number of qubits (100) and linearly increasing number of gates (100, 200, 300, \ldots, 1000); 2) linearly increasing number of qubits (100, 200, 300, \ldots, 1000) and constant number of gates (300). In both scenarios, 50\% of the gate count are T gates.

\begin{figure}[t!]
    \centering
    \includegraphics[width=0.9\columnwidth]{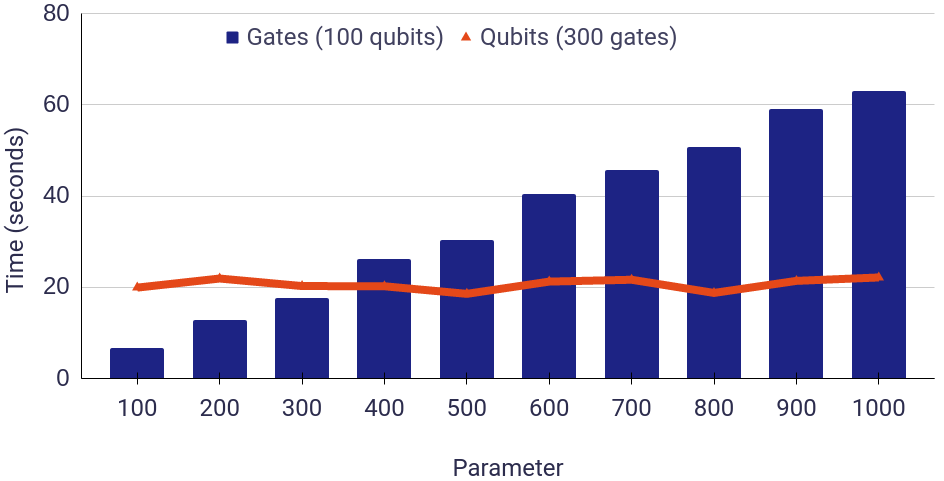}
    \caption{Scalability Benchmark. Compilation time for circuits consisting of: either a fixed number of qubits and increasing number of gates (blue bars), or a fixed number of gates and increasing number of qubits (red line). Compilation time grows linearly with number of gates, and is constant with increasing number of qubits.}
    \label{fig:scalability}
\end{figure}

The gates are translated into multi-body measurements, and these are compiled sequentially. Parallelisation of such measurements is not, for the moment, performed by OpenSurgery. Distillations are executed one at a time, and their scale is set to accommodate two levels. There is a 50\% routing overhead to ensure that all operations can be implemented.

The plot from Fig.~\ref{fig:scalability} illustrates the current scalability of OpenSurgery. Using an i7 processor and 32GB RAM, compilation is under 2 minutes for a circuit of 1000 gates and 100 qubits. The number of qubits does not influence the compilation time, because measurements are not parallelised. 
Implementing parallelisation will be a very complex optimisation problem, which would influence compilation time. Future work will focus on such a feature, because it will reduce the volume of the compiled assemblies. The current version of OpenSurgery generates a constant overhead associated with the A* routing such that the total time overhead for computing routes grows linearly with the number of measurements.

\section{Conclusion}

We introduced OpenSurgery, a tool for preparing topological assemblies using lattice surgery. It is, to the best of our knowledge, the first tool that can manipulate arbitrary circuits, and which has built-in functionality to perform resource estimation for the resulting assembly volumes. OpenSurgery is a first step towards faster design automation of practical quantum algorithms. OpenSurgery includes also a 3D visualisation module. Its scalability was illustrated by preparing large circuits.  Future work will investigate heuristics for compiling more compact assemblies, and methods for improved assembly analytics.

\section*{Acknowledgement}
AP was supported by the Linz Institute of Technology project CHARON, Google Faculty research awards, and a Fulbright Senior Researcher Fellowship.

\bibliographystyle{IEEEtran}
\bibliography{_main}

\end{document}